\documentclass{ws-mpla}
\usepackage[super]{cite}
\usepackage{graphicx}
\begin{document}


\def\beq{\begin{eqnarray}}    
\def\eeq{\end{eqnarray}}      

\newcommand{\Ln}{\,\mbox{Log}\,}
\newcommand{\Det}{\,\mbox{Det}\,}
\newcommand{\Newt}{{\mbox{\tiny Newt}}}
\def\na{\nabla}
\def\pa{\partial}

\def\tr{\,\mbox{tr}\,}                  
\def\rot{\,\mbox{rot}\,}                
\def\sen{\,\mbox{sen}\,}                
\def\Tr{\,\mbox{Tr}\,}                  
\def\Res{\,\mbox{Res}\,}                
\def\Det{\,\mbox{Det}\,}                
\def\Log{\,\mbox{Log}\,}                

\def\al{\alpha}
\def\be{\beta}
\def\ch{\chi}
\def\ga{\gamma}
\def\de{\delta}
\def\ep{\epsilon}
\def\vp{\varepsilon}
\def\ze{\zeta}
\def\io{\iota}
\def\ka{\kappa}
\def\La{\Lambda}
\def\la{\lambda}
\def\ro{\varrho}
\def\si{\sigma}
\def\om{\omega}
\def\ph{\varphi}
\def\ta{\tau}
\def\Om{\Omega}
\def\te{\vartheta}
\def\up{\upsilon}
\def\Ga{\Gamma}
\def\De{\Delta}
\def\La{\Lambda}
\def\Si{\Sigma}
\def\Om{\Omega}
\def\Te{\Omega}
\def\Th{\Theta}
\def\Up{\Upsilon}

\markboth{Authors' Names}
{Instructions for Typing Manuscripts (Paper's Title)}

\catchline{}{}{}{}{}

\title{Gravitational Waves and Perspectives for Quantum Gravity}


\author{Ilya L. Shapiro}

\address{Departamento de F\'{\i}sica, ICE,
Universidade Federal de Juiz de Fora, MG, Brazil
\\
shapiro@fisica.ufjf.br
\\
and
\\
Tomsk State Pedagogical University and Tomsk State University,
Tomsk, Russia}

\author{Ana M. Pelinson}

\address{Departamento de F\'{\i}sica, CFM,
Universidade Federal de Santa Catarina, SC, Brazil\\
ana.pelinson@gmail.com}

\author{Filipe de O. Salles}

\address{Departamento de F\'{\i}sica, ICE,
Universidade Federal de Juiz de Fora, MG, Brazil\\
fsalles@fisica.ufjf.br}

\maketitle

\pub{Received (Day Month Year)}{Revised (Day Month Year)}

\begin{abstract}
Understanding the role of higher derivatives is probably one of
the most relevant questions in quantum gravity theory. Already
at the semiclassical level, when gravity is a classical background
for quantum matter fields, the action of gravity should include
fourth derivative terms to provide renormalizability in the vacuum
sector. The same situation holds in the quantum theory of metric.
At the same time, including the fourth derivative terms means the
presence of massive ghosts, which are gauge-independent massive
states with negative kinetic energy. At both classical and quantum
level such ghosts violate stability and hence the theory becomes
inconsistent. Several approaches to solve this contradiction were
invented and we are proposing one more, which looks simpler than
those what were considered before. We explore the dynamics of the
gravitational waves on the background of classical solutions and
give certain arguments that massive ghosts produce instability
only when they are present as physical particles. At least on
the cosmological background one can observe that if the initial
frequency of the metric perturbations is much smaller than the
mass of the ghost, no instabilities are present.

\end{abstract}

\ccode{PACS Nos.:
04.60.-m, 
11.10.Jj, 
04.30.Nk, 
04.60.Bc, 
}

\section{Introduction}	

General relativity (GR) is a complete theory of classical
gravitational phenomena, which proved valid at the wide range
of energies and distances. However, as any other known physical
theory, it has some limits of application. In order to establish
these limits for GR, one has to review the most important solutions,
which have specific important symmetries.

1) Spherically-symmetric solution is important for describing
objects like planets, stars and black holes.

2) Isotropic and homogeneous metric is used to describe the
zero-order approximation for the Universe.

It is well-known that both of these cases are characterized by
singularities, that means the components of curvature tensor and
energy density of matter become infinite in certain parts of
the space-time manifolds. The natural interpretation is that GR
is not valid at all scales and must be modified in the vicinity
of the singularities. The problem is that, once the action of the
theory and the corresponding equations of motion are modified, these
changes can not be limited to the given regions of space-time.
Indeed, certain modifications can become relevant and in some
cases destructive in all points of the space-time manifolds.
And this is the case for the modifications related to quantum
effects of both matter fields and gravity itself.

One of the most natural reasons to modify GR is related to
quantum effects. The expected scale of the quantum gravity
(QG) effects is associated to the Planck units of length,
time and mass. The three fundamental constants, namely
speed of light $c$, Planck constant ${\hbar}$ and Newton
constant $G$ can be used used uniquely to construct the
universal Planck quantities (we set $c=1$ and ${\hbar}=1$
here)
\beq
l_P = t_P = M^{-1}_P\,,\qquad
M_P =  G^{-1/2} 
\,\approx\, 10^{19}\,GeV\,.
\label{2}
\eeq

One may suppose that the fundamental Planck units indicate to
the presence of a fundamental physics at the Planck scale $M_P$.
Since $c$, ${\hbar}$ and $G$ are involved, it should be relativistic
quantum and gravitational theory at the same time. Unfortunately,
the dimensional approach does not tell us how this unification
should happen. And, as we know, there are plenty of different
ideas of what the relativistic quantum gravitational theory
can look like. Now, in spite of a great variety of approaches,
(more or less) all of them can be classified into three distinct
general groups. One can:

{\bf (i)} \
Quantize both gravity and matter fields. This is the most
fundamental approach and the main subject of the present review,
where we discuss only one particular (albeit very important)
aspect of QG.

{\bf (ii)} \
Quantize only matter fields on classical curved background
(semiclassical approach). There are many good text-books and
introduction reviews on the subject, let us just mention the
books Refs.~\refcite{birdav,GMM,book,ParTom}.
It is important that, different from
quantum theory of metric itself, QFT  and curved space-time are
well-established notions, which passed many experimental and
observational tests. If we put them together, we arrive at the
new physics which is a decent object to explore - just because
it describes some phenomena which certainly exist. On the other
hand, there is no absolute certainty that metric itself should
be quantized, since metric is, after all, different from all
other fields.

{\bf (iii)} \
Instead of quantizing gravity and/or matter, one can quantize
something else. For example, in the case of (super)string theory
both matter and gravity are induced. In this sense string theory
is the most complete and consistent version of QG. However, there
is still a fundamental question of why this theory should be
regarded as quantum gravity. In other words, string theory is
a very important approach to QG, but still it is one possible
alternative among many possible approaches to QG.

An important detail is that all three possible lines of thought
about QG have one common point, namely in all cases one meets
higher derivatives in the effective action of gravity. In
particular, the terms
\beq
S_{HD}=\int d^4x\sqrt{-g}\left\{
\al_1 R_{\mu\nu\al\be}^2 + \al_2 R_{\al\be}^2 + \al_3 R^2
+ \al_4 {\Box}R \right\}
\label{HD-RRR}
\eeq
actually emerge in all three approaches.
In case of string theory, these higher derivative terms
can be reduced and made non-offensive by means of the Zweibach
reparametrization \cite{Zwei}. Indeed, one can perfectly well
use the same approach in the
first two cases, of {\bf (i)} and {\bf (ii)}. It is sufficient
to assume that, after the effective action of gravity
(effective action of the external metric field, in case of
{\bf (ii)}) is found, one can perform the reparametrization
of the metric
\beq
g_{\mu\nu} \longrightarrow
g'_{\mu\nu} =
g_{\mu\nu} +
x_1\,R_{\mu\nu} + x_2\,R\, g_{\mu\nu}\, + \, ...\,,
\label{repar}
\eeq
where the coefficients $x_{1,2,...}$ are chosen in such a
way that the higher derivative terms do not contribute to the
propagator of gravitational perturbations. After that we
define that the corresponding metric $g'_{\mu\nu}$ is
physical, exactly because in this parametrization there
are no dangerous ghosts which can produce instabilities.
In case of the fourth-derivative theory it is sufficient
to require that the first
two terms in (\ref{HD-RRR}), after the transformation
(\ref{repar}), form the combination $\al_1 \big(R_{\mu\nu\al\be}^2
- 4 R_{\al\be}^2\big)$, like in the Gauss-Bonnet topological term.
In this case there is no gauge-independent tensor ghost in
the spectrum. The coefficient of the last term, $R^2$, can
not defined by the requirement of the absence of ghost, hence
it represents an ambiguity. This is a particular case of the
general situation, related to a serious ambiguity in the
physical predictions of the gravitational theory based on
the  transformation (\ref{repar}) \cite{marot}.

In the cases of  {\bf (ii)} and {\bf (iii)} such a
reparametrization concerns only external field (metric)
and hence does not violate the unitarity of the $S$-matrix
for quantum fields.
As far as we know, the phenomenological consequences of this
choice of the metric were never explored in the framework
of the semiclassical approach {\bf (ii)} and we will not
deal with this problem in this review neither. Instead, we
shall assume that no reparametrization like (\ref{repar})
is performed and discuss the effect of ghosts on the
stability of the classical solutions.

In the next part of this short review we shall describe why higher
derivatives are necessary in both semiclassical and quantum
gravity. Furthermore, we present a brief review of ghosts and
of the known approaches to avoid the inconsistencies related to
them.
Finally, we shall give a brief qualitative exposition of the
results of our recent works \cite{GW-Stab,HD-Stab} and show that
there are certain chances that the situation with ghosts is
actually more optimistic when one is dealing with the physics
of QG below the Planck scale.

The review is organized as follows. The next Section is devoted
to the general status of higher derivatives in gravity. In Sect.
3 we consider the known methods of dealing with the ghost problem.
Sect. 4 contains the main results on the effective approach to
such ghosts and, finally, in Sect. 5 we draw our Conclusions.

\section{Why higher derivatives?}	

Let us consider why higher derivatives are necessary in
semiclassical and quantum gravity theories and what the
ghost means.

\subsection{Semiclassical gravity}

The QFT in curved space requires introducing a generalized action
of external gravity field. One can prove that the theory can be
renormalizable only if such a vacuum action includes four derivative
terms. We will not go into details, but just refer the reader to
the books in Refs.~\refcite{birdav}, \refcite{book} for a general
introduction, and to the recent paper in Ref.~\refcite{RenCurved}
for the most complete proof, including the case when non-covariant
gauge fixing conditions are used.
The necessary form of the ``vacuum action'' is as follows:
\beq
S_{vac} &=& S_{EH}\,+\,S_{HD}\,,
\label{vacuum}
\eeq
where
\beq
S_{EH} &=& -\,\frac{1}{16\pi G}\,
\int d^4x\sqrt{-g}\,\left\{R + 2\La \,\right\}\,
\label{EH}
\eeq
is the Einstein-Hilbert action with the cosmological constant
and the higher derivative term $S_{HD}$, defined in (\ref{HD-RRR}),
can be recast in the most useful form as
\beq
S_{HD} &=& \int d^4x\sqrt{-g}\left\{
a_1C^2+a_2E+a_3{\Box}R+a_4R^2\right\}\,.
\label{HD}
\eeq
Here
\beq
C^2 &=& R_{\mu\nu\al\be}^2 - 2R_{\al\be}^2 + 1/3\,R^2
\label{W}
\eeq
is the square of the Weyl tensor and
\beq
E &=& R_{\mu\nu\al\be}R^{\mu\nu\al\be}
-4 \,R_{\al\be}R^{\al\be} + R^2
\label{GB}
\eeq
is the integrand of the topological Gauss-Bonnet term.

\begin{figure}
\centerline{\includegraphics[width=2.4in]{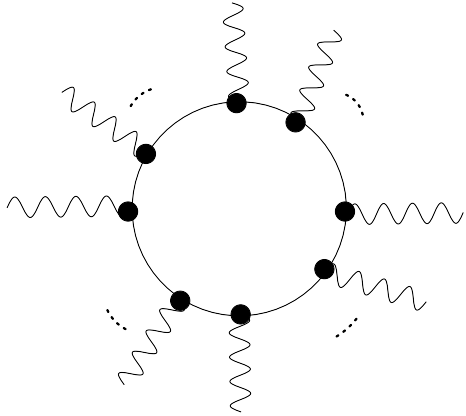}}
\caption{General one-loop diagram of matter loop with gravitational
external lines. The number of such lines coming to each vertex can
be arbitrary. The divergences emerge only from the diagrams with
one and two gravitational vertices. The power counting does not
change at higher-loop order, if the matter theory is flat-space
renormalizable.}
\label{ScalVec}
\end{figure}

In order to understand why the terms (\ref{vacuum}) are necessary,
it is sufficient to consider the one-loop approximation, when the
vacuum contribution reduce to the single bubble of matter field,
as shown in Fig. \ref{ScalVec}.
The external gravity can be implemented by means of the external
tails of the field $h_{\mu\nu}$, where we assume the linear
parametrization of the metric, $g_{\mu\nu}=\eta_{\mu\nu}+h_{\mu\nu}$.
The simple bubble without tails of $h_{\mu\nu}$ has quartic
divergences, the bubble with one vertex of matter-matter-$h_{\mu\nu}$
interaction has quadratic divergences and, finally, the bubble with
two such vertices has logarithmic divergences. Due to the general
covariance of the counterterms \cite{RenCurved}, the possible
logarithmically divergent structures are exactly those listed in
(\ref{vacuum}).

In quantum gravity, higher derivative term like the square of
the Weyl tensor indicate the presence of massive ghost, namely,
a spin-two particle with negative kinetic energy. This leads to
the problem with unitarity, at least at the tree level. But, in
the semiclassical theory, gravity is external and unitarity of
the gravitational $S$-matrix can be not requested. Therefore,
the consistency conditions in this case can be relaxed to the
following:
\ (i) existence of physically reasonable solutions and
\ (ii) their stability under small metric perturbations.
Perhaps the most important point here is that the theory without
the fourth-derivative terms (\ref{HD}) can not be consistent. If
we do not include them into the classical action, these terms will
emerge in the quantum corrections anyway, with infinite coefficients.
The difference with the theory (\ref{vacuum}) would be that, without
these terms in the classical action one can not control higher
derivatives by means of multiplicative renormalization.

The last two important observations concerning higher derivatives
are as follows. First, if we consider a more general theory, where
metric is also an object of quantization, anyway one should
definitely quantize both matter and gravity, for otherwise
the QG theory would not be complete. Then the diagrams with
matter internal lines in a complete QG are be exactly the same
as in a semiclassical theory. This means one can not quantize
metric without higher derivative terms in a consistent manner,
since these terms are produced already within the semiclassical
theory. The second point is that the effect of higher derivative
terms can be Planck-suppressed in the classical solution, but
this does not solve the problem, because the canonical wisdom
tells us that more massive ghost would be even more destructive
and lead to even faster explosion of the vacuum space-time
(see, e.g., Ref.~\refcite{Woodard-07} for discussion). As we
shall see in what follows, this is not necessary the case, so
the huge mass of the ghost can provide a solution of the whole
problem.


\subsection{(Super)renormalizable quantum gravity}

Let us now briefly consider the situation with higher derivatives
in a proper QG. The standard traditional approach to QG assumes
that the quantization can be performed using the variables
\beq
\ka h_{\mu\nu}=g_{\mu\nu}-\eta_{\mu\nu}\,.
\label{2201}
\eeq
It is assumed that this choice of the quantum metric is
the ``right'' one. This choice enables one to deal with the
well-defined object such as $S$ - matrix of the gravitational
field $h_{\mu\nu}$. Let us note that in the case of quantum
General Relativity the parametrization (\ref{2201}) corresponds
to so-called Gaussian expansion of the action
\beq
S_{EH}\,=\,-\,\frac{1}{\ka^2}\,
\int d^4x\sqrt{-g}\,R\,,
\qquad
\ka^2 = 16\pi G\,.
\label{2202}
\eeq
This means that in the variables (\ref{2201}) the second order
(Gaussian approximation) of the action is $\ka$-independent.
In other words, in the absence of cosmological constant, the
expansion in the coupling $\ka$ and the flat-space expansion
(\ref{2201}) are related notions. One of the consequences of
this is that in the theories with a non-Gaussian fixed point
$\ka_0$ \cite{Assafe-1,Assafe-2}, the natural expansion is not
the one in the powers of $\ka$, but in the powers of $\ka-\ka_0$.
In this case, the flat space is not a distinguished space-time
from quantum viewpoint. On the other hand, this relation shows
that the $S$-matrix approach to QG is relevant mainly in the
standard perturbative QG. In the non-perturbative approaches,
such as asymptotic safety scenarios, there may be no well-defined
asymptotic states, no well-defined $S$-matrix and, hence, one
should look for another criterion for the consistency of the
theory, without necessary using the unitarity of the $S$-matrix
as such a criterion. Once again, we arrive at the idea of checking
stability of the physically relevant classical solutions in the
theory. One can certainly consider this to be a minimal condition
of consistency of the theory.

The construction of QG starts from some covariant action of gravity,
\beq
S &=& \int d^4x \sqrt{-g}\,\,{\cal L}(g_{\mu\nu})\,,
\label{act}
\eeq
where
${\cal L}(g_{\mu\nu})$ is the covariant Lagrangian density. The
action can be (\ref{EH}), or  (\ref{vacuum}), or some other, with
more higher derivative terms. The gauge transformation is the
diffeomorphism ${x^\prime}^\mu = x^\mu + \xi^\mu$ and the metric
transforms under it as
\beq
\de g_{\mu\nu} &=& g^\prime_{\mu\nu}(x) - g_{\mu\nu}(x)
= - \na_\mu\xi_\nu - \na_\nu\xi_\mu\,.
\label{met}
\eeq
By using the method described in Ref.~\refcite{Stelle77} for the case of
(\ref{vacuum}), one can prove that the effective action of the
metric, $\Ga(g_{\mu\nu})$ is also  diffeomorphism invariant and
the possible divergences are local, as usual in quantum field
theory. Then one can use the notion of power counting to explore
the possible form of these divergences.

The universal formula for the superficial degree of divergence
of a diagram with the power of momenta $r_l$ is the inverse of
the propagator of some internal line, with the number of vertices
$n$ and $K_\nu$ power of momenta in a given vertex, is
\beq
D + d \,=\,\sum\limits_{l_{int}}(4-r_l)
\,-\,4n \,+\,4\,+\,\sum\limits_{\nu}K_\nu\,.
\label{index}
\eeq
Here $D$ is the superficial degree of divergence of a diagram and
$d$ is the total number of derivatives on its external lines.
Furthermore, one can use topological relations, e.g., the one
between number of loops $p$, vertices $n$, and internal lines
\beq
l_{int} \,=\, p + n - 1\,.
\label{top}
\eeq

As the first example, let us consider quantum GR, with the action
(\ref{EH}). Obviously, the diagrams with the vertices $K_\nu=0$ will
be less divergent. Therefore, for the sake of simplicity we consider
only vertices with $K_\nu=2$. Taking $r_l=K_\nu=2$ and combining
eqs. (\ref{index}) with (\ref{top}), we arrive at the estimate
\beq
D+d\,=\,2+2p\,.
\label{dD}
\eeq
One can see that the diagrams with $D=0$ (that means only logarithmic
divergences) require counterterms with the number of derivatives of
the metric growing with the loop order according to $d=2+2p$. This
means that quantum GR is not renormalizable. This conclusion has
been supported by direct calculations at one \cite{hove,dene} and
two \cite{gosa} loops. In the last case there is the counterterm
$R_{\mu\nu\al\be}R^{\mu\nu}\,_{\rho\si} R^{\mu\nu\rho\si}$, which
does not vanish on-shell, that means the divergences of the
$S$-matrix of the theory can not be dealt with in a regular way.
In the presence of matter the situation is similar even at the
one-loop level.

Within the standard perturbative approach non-renormalizability
means the theory has no predictive power. Every time we introduce
a new type of counterterms, it is necessary to fix renormalization
condition and this means performing a measurement. Since,
according to (\ref{dD}), the number of counterterms is not
restricted, before making a single prediction, in principle
it is necessary to have an infinite amount of experimental data.

There are two possible way to the solutions of this problem:

1. Change standard perturbative approach to something else.
This is the line of research which is beyond the framework of
the present review. Let us only say that there are many options
in this area, but their consistency and relation to the standard
QG program are not clear, in all cases.

2. One can change to another theory as a starting point to
construct QG.
\vskip 1mm

The first option is widely explored in the asymptotic
safety scenarios, in the effective approaches to QG and in
the induced gravity paradigm (including string theory).
Let us concentrate on the second choice, that means we will try to
change the theory which is the subject of quantization. In this
case one meets a great variety
of different models. For example, the string theory belongs
also to this class of approaches. However, if one is looking for
a most simple and natural solution, the first option should be
just to introduce higher derivatives into the gravitational
action, starting from the terms (\ref{HD}). The reason is that
we need these terms anyway for quantizing matter field. Since
the full QG theory should include both metric and matter being
quantized, one has to take care about the diagrams with the
internal lines of matter fields and external ones of the
gravitational field $h_{\mu\nu}$. But those diagrams are exactly
the ones of the semiclassical theory and, as we already noted,
they require the terms of both  (\ref{HD}) and  (\ref{EH}) to
be included.

Now, if we do so, the situation in QG changes dramatically, because
this new theory is renormalizable. The propagators and vertices in
higher derivative QG (HDQG) are not the same as in quantum GR.
In this case we have $r_l=4$ and there are vertices with
$K_4$, $K_2$ and $K_0$. Then the superficial degree of divergence
can be easily evaluated to give
\beq
D+d &=& 4-2K_2-2K_0\,.
\label{hd}
\eeq
This theory is definitely renormalizable. Dimensions of counterterms,
at all loops, are $4,\,2,\,0$, depending on the number of the
vertices of the $K_2$- and $K_0$-type.

Unfortunately, there is a high price to pay for renormalizability.
The higher derivative QG based on the fourth-derivative action
(\ref{vacuum}), possesses a massive spin-two
gauge-independent excitation called massive ghost,
\beq
G_{\rm spin-2}(k)\,\sim\,\frac{1}{m^2}\,\,\left(
\frac{1}{k^2} - \frac{1}{k^2+m^2} \right),\quad
m \propto M_P\,.
\label{ghost}
\eeq
In the framework of linearized theory one can separate the massless
and massive degrees of freedom. It is an easy exercise to check that
the kinetic energy of the massive component is negative. For this
reason this particle is called massive ghost. Indeed, the mass of
this ghost is huge, of the Planck order of magnitude. The main point
of this review is a new proposal concerning ghosts, which was
originally done in Ref.~\refcite{HD-Stab} (see also Ref.~\refcite{GW-Stab}).

Including even more derivatives was initially thought to move massive
pole to a much higher mass scale. In Ref.~\refcite{highderi} the
following action was proposed
\beq
S  &=&  S_{EH} \,+\,\int d^4x\sqrt{-g}\,\Big\{
a_1R_{\mu\nu\al\be}^2 + a_2R_{\mu\nu}^2 + a_3R^2 + \, ...
\nonumber
\\
&+&
c_1 R_{\mu\nu\al\be} \Box^k R^{\mu\nu\al\be}
+ c_2 R_{\mu\nu}\Box^k R^{\mu\nu}
+ c_3 R\Box^k R
+ \, b_{1,2,..}R_{...}^{k+1}
\Big\}\,.
\label{highde}
\eeq
A simple analysis shows that this theory is superrenormalizable,
but the massive ghost is still here. For the case of real poles
one can prove that the spin-two part of the propagator has the
structure
\beq
G_2(k) &=& \frac{A_{0}}{k^2} +
 \frac{A_{1}}{k^2 + m_1^2} + \frac{A_{2}}{k^2 + m_2^2}
+ \cdots + \frac{A_{N+1}}{k^2 + m_{N+1}^2}\,,
\label{propa}
\eeq
where for any sequence $0 < m_1^2 < m_2^2 < m_3^2 < \cdots < m_{N+1}^2$,
the signs of the corresponding terms alternate, $A_j\cdot A_{j+1} < 0$.
Therefore, the situation when the ghost is shifted to an infinite
energy level is ruled out.

In order to understand the renormalizability properties of the theory
one has to consider the power counting. It is sufficient to consider
only vertices $K_\nu = 2k+4$, which produce strongest divergences.
Then we have $r_l=K_\nu=2k+4$ and one can easily arrive at the estimate
\beq
D+d &=& 4\,+\,k(1-p)\,.
\label{indi-HD}
\eeq
For $k=0$ we meet the standard result, $d\equiv 4$ for $D=0$.
Starting from $k=1$ we have superrenormalizable theory,
where the divergences exist only for $p=1,2,3$.
For $k\geq 3$ we have superrenormalizable theory, where
the divergences exist only at the one-loop level, when $p=1$.

So, if we solve the ghost problem someday, there will be many
versions of renormalizable and superrenormalizable higher
derivative QG theories. In particular, superrenormalizable QG theory
described above has some interesting features.
For example, in the basic fourth-derivative version, the quantum
corrections have an ambiguity related to the choice of gauge-fixing
condition. As a consequence of this, there are no well-defined
$\be$ - function for the Newton constant $G$ and for the cosmological
constant $\Lambda$. On the contrary, in the superrenormalizable
versions there is no such problem, in fact all $\be$  - functions
are well-defined and gauge-fixing independent. Moreover, for
$k \geq 3$ it is possible (albeit very difficult)
to derive exact $\be$-functions. Finally, one can say that there
are quite a lot of superrenormalizable QG theories with many free
parameters, and not a single experiment to fix their values.
Perhaps, the main problem of QG is not just a theory,
but the lack of experimental data.

\section{What is the problem with ghosts?}

A consistent theory of quantum matter fields on classical curved
background can be achieved only if we include the higher derivative
terms (\ref{HD}) into the classical action of vacuum. The same
action (\ref{vacuum}) represents also a basis for renormalizable
QG theory. What is a problem with massive ghosts which make
the higher derivative theories so problematic?

In short, the presence of ghosts created the following problem:
the vacuum state of the theory becomes unstable and theory
gets inconsistent. This means, for example, that the vacuum is
not protected against a spontaneous creation of massive ghost and
of a certain amount of normal particles, which compensate the
negative energy of the ghost. Other forms of formulating the
same problem can be formulated as follows:
\vskip 1mm

In classical systems higher derivatives generate exploding
instabilities at the non-linear level, as was discussed
originally by Ostrogradsky in 1850 in Ref.~\refcite{Ostrogradski}
and in relation to the gravitational case by Woodard
in Ref.~\refcite{Woodard-07}. One can check that at the linear
level the theory is stable, therefore the problem is related to
the non-linear level of consideration.
\vskip 1mm

Interaction between ghost and gravitons may violate energy
conservation in the massless sector. This possibility has
been explored by Veltman in 1963 \cite{Veltman}.
In short, this means that the quantum scattering of ghost
and normal (positive-energy) particles has the feature that
the dominating process is accelerating a ghost, which gains
more and more negative energy, while the positive compensating
energy goes to the outflux of gravitons, in the present case.
\vskip 1mm

So, the situation is such that the presence of ghost makes the
theory ill-defined. Therefore, the question is whether it is
possible or not that the Lagrangian of the theory admits the
existence of the massive ghost excitation, while there is no
such particle in reality. The first observation is that, if we
do not include the ghost into the {\it in} state, the theory
will be inconsistent, because interactions between ghost and
gravitons will produce the ghost particles in the  {\it out}
state and the $S$-matrix of the theory will be non-unitary.
Indeed, one can think that the scattering problems in QG are
not the most important ones, especially at the relatively
low energies, which means just much below the Planck scale.
Assuming this viewpoint, one may not care about the $S$-matrix,
especially because the Planck-mass particle may be
non-observable. However, this would be a wrong conclusion,
because if massive ghosts really emerge in the theory, this
will result in the Veltman scattering and finally to the huge
flux of the energy of gravitons. Since we do not observe the
emergence of gravitons with the Planck energy densities in
experiments (and, in fact, in the real life which would be
seriously affected!), it is natural to conclude that the
Nature is organized somehow different, such that we do not
have the Planck-scale ghosts interacting to gravitons.

\section{History of the fight: physicists against ghost}

Hopefully we have convinced the reader that the situation is
somehow contradictory. From one side, we need higher derivatives
to construct renormalizable theory of matter fields and quantum
theory of the metric itself. Up to some extent the first part is
even more relevant, because one can deal with the quantization
of the metric in different ways, e.g., switch to the string theory.
However, in the case of semiclassical theory we have to deal
with a real physics, that means to put quantum fields on curved
background. And this really requires the terms (\ref{HD}) to be
present in the action. On the other hand, such important
achievement as Hawking radiation is, \cite{Hawking75} in fact,
related to the conformal anomaly,\cite{ChrFull} which results
from the renormalization of the terms (\ref{HD}). The same is
true for another very important application of quantum theory.
The complete version of the Starobinsky model of inflation
\cite{star} is also based on the conformal anomaly \cite{fhh}.
In both cases one can perform the analysis on the basis of the
anomaly-induced effective action of gravity \cite{rie}, which
is just a natural quantum extension of (\ref{HD}). So, the
higher derivatives are necessary and important, hence we can
not (and, in some sense, do not like to) get rid of them.

On the other hand, in the presence of higher derivatives one
meets massive ghosts, vacuum instability and related difficulties
with consistent formulation of the theory at both classical and
quantum levels. Hence, it is not a surprise that there were many
attempts to find a solution of the ghost problem. Let us briefly
discuss some of them.

One can construct higher derivative
theories of gravity without ghosts \cite{Neville,seznie} (see
also recent works in Ref.~\refcite{Hela}). All these theories are
very similar to the Gauss-Bonnet action, which is free
of ghosts because the propagator of the gravitational field
behaves in the UV like $1/k^2$. As it was discussed in
Ref.~\refcite{torsi}, this rules out renormalizability in the QG
theory. The situation is even more clear in the semiclassical
case, when the terms of the action (\ref{HD}) are generated
by the matter loops. Obviously, any modification of the
gravity action which rules out these terms does not help to
make the theory consistent.

The mainstream approach for the ghost problem has been
developed in Refs.~\refcite{Tomboulis,SalStr,AntTom}.
The idea is to assume the resummation of the perturbative
series such that, taking the full propagator instead
of the tree-level one, massive ghost becomes unstable
and disappears in the {\it out} state. Then one can start
with the {\it in}-state without ghosts and still have an
unitary $S$-matrix of the gravitational perturbations.
From the technical side, the best realization of this idea
requires that the loop corrections shift the
position of the ghost pole to the complex plane and that this
position becomes gauge-fixing dependent.\cite{ AntTom}  With these
assumptions one can prove the unitarity of the theory.
Unfortunately, the existing perturbative and non-perturbative
(e.g., $1/N$ expansion, lattice formulations etc) methods are
not sufficient to claim whether these two conditions or
at least part of them are satisfied or not \cite{Johnston}.

An alternative idea has been suggested by Hawking et al
\cite{Hawking}. It is
necessary to remember that the ghost does not emerge in the
theory as independent particle, but only comes together with
graviton. There is a chance that the quantum field theory which
takes this aspect into account, will be free of instabilities.
The realization of this interesting idea requires qualitatively
new formulation of quantum field theory. Unfortunately, until
now there is no consistent formulation of this sort and hence
one can not be sure whether this approach really works or not.

Another possibility is to consider the situation when the
ghosts are present, but the  decay of vacuum takes very long
time \cite{Antoniad,GarVil}. There is an explicit calculation
supporting this possibility \cite{Maggiore-13}, but it works
only for a very small mass of the ghost. On the other hand,
in case of higher derivative quantum gravity the typical
mass of the ghost is very large, of the Planck order of
magnitude.	

Finally, there was a very interesting proposal to generalize
(\ref{highde}) to the non-polynomial in $\Box$ structure
\cite{Tomb97}. One can choose the function of $\Box$ in
such a way that there would be no ghosts. It is supposed that
this theory will be superrenormalizable, exactly as (\ref{highde})
is for the polynomial of high (above three) order function.
Recently there was a significant activity in exploring this
kind of theories at the classical level (see
Ref.~\refcite{Modesto} for the review and further references).
At the quantum level, the main questions are how to perform
quantization in the non-polynomial theory, and how to perform
practical calculations. Another difficult problem is how to
evaluate the superficial degree of divergence in such a
theory. Gravity has a non-polynomical interactions, and
according to the general formula (\ref{index}) we meet an
indefinite output $D+d \sim \infty - \infty$. Then it is
difficult to arrive at some definite conclusions about whether
the theory under discussion is superrenormalizable,
renormalizable or even non-renormalizable. Some relevant
considerations about this subject can be found in
Ref.~\refcite{Modesto-2}.

We can conclude that the situation with Planck-scale massive
ghosts is unclear, in the sense there are several interesting
proposals, but no certainty that al least one of them can be
successful. In what follows we describe a new approach which
is much simpler and is probably working. The main point is the
possibility of that the ghosts may be actually not generated
from vacuum by interaction with gravitons, if these gravitons
do not have energies comparable to the Planck scale. One can
note that the spontaneous creation of ghost and gravitons means
that the energy density of these gravitons in the given
space-time point is of the Planck order of magnitude.

The low-energy classical solutions of the theory (\ref{vacuum})
with higher derivative terms (\ref{HD}) should be very close
to the ones of GR, because the effect of the  terms (\ref{HD})
is going to be Planck-suppressed. Then the theory is, in general,
safe from the dangerous ghost instabilities if these classical
solutions are stable against small perturbations of the metric.
The stability of the theory in the Lyapunov sense (see, e.g.,
Ref.~\refcite{Math}) means that the non-trivial background takes care
about all non-linear effects, so if the theory is stable with
respect to the small perturbations at the linear level, then
the non-linear stability is guaranteed. On the other side, it
is known that the spontaneous creation of ghost does not occurs
within the linearized theory in flat space-time background (see,
e.g., Ref.~\refcite{Woodard-07}).  Therefore, all the question
is whether the background metric can change this situation
and, if this is the case, what are the conditions for the
creation of ghost and/or instabilities.

\section{Stability of classical solutions at low energies}

In our opinion, the most risky assumption which is usually done
to rule out the higher derivative theory is that the Ostrogradsky
instabilities or Veltman scattering are relevant independent on
the energy scale. There is a relatively simple way to check this
assumption. Let us take a higher derivative theory of gravity and
verify the stability with respect to the linear perturbations on
some, physically interesting, dynamical background. If the
mentioned assumption is correct, we will observe rapidly growing
modes even for the low-energy background and for the low
initial frequencies of the gravitational perturbation.
On the contrary, if there are no growing modes at the linear
level, there will not be such modes even at higher orders.
One has to remember that the ghost issue is essentially a
tree-level problem, so the study of classical solution is
sufficient to draw conclusions about the general situation.

Up to the present moment, the program formulated above has
been realized in the following three cases:
\vskip 1mm

1) Cosmological background. In the particular case of de Sitter
metric the result is partially known for more than thirty years
\cite{star83}
and has been repeatedly confirmed \cite{wave,HHR}. In these papers
the theory with semiclassical corrections to the classical action
(\ref{HD}) has been used. On the other hand, recently the same
investigation has been repeated for other cosmological metrics,
such as radiation- and dust-dominated Universes \cite{GW-Stab}.
At the same time, in all these papers the relation between
instabilities and higher derivative ghosts was never traced back
explicitly. This last part has been explored in Ref.~\refcite{HD-Stab}
and in what follows we shall review the main results of this
work.
\vskip 1mm

2) Black hole background. In this case there are conflicting
data in the literature, namely the statements about
stability \cite{Whitt} and instability \cite{Myung} of this
solution. The analysis of this case is technically very
complicated and we will not discuss it here in details. Let us
only mention that it is not clear, to which extent the results
depend on the choice of the boundary conditions, on the
frequency of initial seeds of perturbations and also on some
technical assumptions done in these works.
\vskip 1mm

3) General curved background which is close to a flat space-time.
Since the non-linearities of the perturbations can be taken into
account by means of a non-trivial metric background, it looks
natural to consider a weak (albeit arbitrary) gravitational field.
Such consideration is, in principle, possible \cite{HD-Stab}.
Using normal coordinates and local momentum representation, we
have constructed the relevant equations in the lowest non-trivial
approximation. It is natural to expect that there will be a smooth
transition to the precisely flat case, where ghost does not lead
to instability. Probably this, technically complicated, study will
end with the conclusion that there are no instabilities if the
curvature is much smaller than the square of the Planck mass,
which is the unique dimensional parameter in the theory
(\ref{vacuum}). However, the results are not conclusive yet in
this part.
\vskip 1mm

Taking the present-day state of art into account, in what follows
we shall describe only the situation with the cosmological
background, where the results are qualitatively clear and
understandable.

\section{Background cosmological solutions}

In principle, one can explore the stability of the classical
solution in the theory (\ref{vacuum}), but for the sake of
generality we can include also the semiclassical corrections
coming from the massless fields. It is supposed that the effects
of massive fields are negligible at the sufficiently low
energies.

In the case of massless conformal fields one can set to zero the
coefficient of the $R^2$-term, $a_4=0$, in the action (\ref{HD}),
without violating renormalizability. Therefore, the theory of
our interest is described by the sum of a classical action
(\ref{vacuum}) with $a_4=0$ and with the additional anomaly-induced
quantum contribution\footnote{Since we are mainly interested
in the tensor gauge-independent mode of the metric perturbations,
there is no problem to assume that the classical $R^2$-term
is absent, since this term does not influence too much the
dynamics of this mode.},
\beq
\Gamma_{ind} &=& S_c[g_{\mu\nu}]
\,-\, \frac{3c+2b}{36(4\pi)^2}
\,\int\limits_x\,R^2(x)
\,+\,
\frac{\om}{4}\,\iint\limits_{x\,y}\,
C^2(x)\,G(x,y)\,\big(E - \frac23{\Box}R\big)_y
\nonumber
\\
&+& \frac{b}{8}\,
\iint\limits_{x\,y}\,
\big(E - \frac23{\Box}R\big)_x\,G(x,y)\,
\big(E - \frac23{\Box}R\big)_y\,,
\label{induc}
\eeq
where we used compact notations
\beq
\int_x =\int d^4 x\sqrt{-g}\,,\quad
\mbox{and}
\quad
\De_4\,G(x,y)=\de(x,y)\,.
\label{integ}
\eeq
Furthermore,
\beq
\Delta = {\Box}^2 + 2R^{\mu\nu}\na_\mu\na_\nu
- \frac23\,R{\Box} + \frac13\,(\na^\mu R)\na_\mu
\eeq
is the conformal self-adjoint Paneitz operator, coefficients
$\om,\,b,\,c$ depend on the number of quantum fields and
$S_c[g_{\mu\nu}]$ is an arbitrary conformal invariant functional
of the metric. Further details about derivation of (\ref{induc})
can be found, e.g., in Ref.~\refcite{wave}.

In order to understand the effect of quantum terms on the
conformal factor of the metric, let us consider the equation
for this factor $a(t)$,
Consider unstable inflation, matter
(or radiation)
dominated Universe and assume that the Universe is close
to the classical FRW solution. The equation is
\beq
\frac{{\ddot {\ddot a} }}{a}
+\frac{3{\dot a} {\dot {\ddot a}} }{a^2}
+\frac{{\ddot a}^2}{a^2}
-\left( 5+\frac{4b}{c}\right)
\frac{{\ddot a}{\dot a}^2}{a^3}
-2k\left( 1+\frac{2b}{c}\right)
 \frac{{\ddot a}}{a^{3}}
\nonumber
\\
-\frac{M_{P}^{2}}{8\pi c}
\left( \frac{{\ddot a}}{a}+
\frac{{\dot a}^2}{a^{2}}
+\frac{k}{a^{2}}-\frac{2\Lambda }{3}\right)
\,=\,-\,\frac{1}{3c}\,\rho_{matter}\, ,
\label{equa}
\eeq
where we have also introduced the matter term for illustrative
purpose. Also, $k=0,\pm 1$ and \ $\La $ is the cosmological constant.
It is easy to see how the things change in this equation
when the time change. First of all, let us consider the empty
universe, with $\rho_{matter} \to 0$. In this case one can
find particular solutions  (see Ref.~\refcite{star} and also Ref.~\refcite{asta}
for the case with cosmological constant)
\beq
a(t) \,=\,  \left\{
\begin{array}{c}
a_o\,\exp(Ht)\,,\qquad k=0 \,\,\,\,
\\
a_o\,{\rm cosh}(Ht)\,,\qquad k=1 \,\,\,\,\,\,
\\
a_o\,{\rm sinh}(Ht)\,,\qquad k=-1
\end{array}\right.\,,
\eeq
where Hubble parameter takes two constant values
\beq
H\,=\,H_{\pm}\,=\, \frac{M_P}{\sqrt{-32\pi b}} \, \left(\,1\pm \,
\sqrt{1+\frac{64\pi b}{3}\,\frac{\Lambda }{M_P^2}
}\,\,\right)^{1/2}\,.
\label{pm}
\eeq
Let us note that the coefficient $b$ is negative for any
particle content of the theory contributing to quantum terms.
For the small cosmological constant $\La \ll M_P^2$ the two
solutions (\ref{pm}) boil down to
\beq
H_{+}\,=\, \frac{M_P}{\sqrt{-16\pi b}}
\,,\qquad
H_{-}\,=\, \sqrt{\frac{\La}{3}}\,.
\label{pm boil}
\eeq
Obviously, the first solution here is usual Starobinsky
inflation (initial part of it, better say) and the second
one is the usual dS solution without quantum corrections.
What we need here is the stability of the second of these
solutions with respect to the tensor perturbations of
the metric.

One can first perform a simple test of the model, by considering
the stability of the low-energy solution with $H_{-}$ with
respect to the perturbations of the conformal factor
(see second reference in Ref.~\refcite{asta}).
Consider $H \to H_{-} + const \cdot e^{\la t}$ and arrive at
\beq
\la^3+7H_0\la^2 + \left[\frac{(3c-b)4{H_0}^2 }{c}
- \frac{M^2_P}{8\pi c}\right]\la
\,- \,\frac{32\pi b{H_0}^3+M^2_PH_0}{2\pi c}\,=\,0\,.
\eeq
The solutions of this equation have positive real parts
\beq
\la_1=-4H_0\,,\qquad
\la_{2/3}\,
=\,-\frac32\,H_0\,\pm\,\frac{M_P}{\sqrt{8\pi |c|}}\,i\,,
\eeq
indicating the absence of growing modes.
Obviously, the positive cosmological constant $\La>0$
protects the low-energy dS solution from
higher-derivative instabilities in this case.

One can regard the two dS solutions (\ref{pm boil}) as extreme
states of the Universe \cite{Shocom,asta}. The first of these
solutions is the
initial phase of the Starobinsky inflation and the last one is the
distant future of the Universe when the effect of all kinds of
matter becomes irrelevant and only cosmological constant will
drive the accelerated expansion. What is important for us is
that, in the low-energy regime of a late Universe, the solution
with $H_{-}$ provides an extremely precise approximation
for the solution with quantum terms taken into account. In the
absence of quantum term (\ref{induc}) this is an exact solution,
because Eq. (\ref{HD}) with $a_4=0$ does not affect the dynamics
of the conformal factor. But even if the quantum term (\ref{equa})
is taken into account, it is still a perfect approximation. The
reason that the theory without matter has only two dimensional
parameters, $M_P$ and $\La$. Any correction to $H_{-}$ is given
by a positive power of the ratio $\La/M_P^2$, which is of the
order of $10^{-120}$. So, we can safely use this background
solution at low energies.

Let us now consider the case with the nontrivial matter
contents, $\rho_{matter}$. Consider the late time epoch.
It is easy to see that the terms of the first line of
(\ref{equa}), which are of the quantum origin, behave like
$t^{-4}$. At the same time the second-line terms, of the
classical origin, all behave like $t^{-2}$ (see Ref.~\refcite{radiana}
for more detailed discussion). Obviously, the quality of the
classical approximation for the solution $a(t)$ becomes
better for $t \to \infty$ and can be considered a very good
one in the late epoch of the Universe.

\section{Gravitational waves and ghosts}

Now we are in a position to explore the dynamics of the gravitational
waves on the background of cosmological solutions described in the
previous section. For this end we consider small perturbation
\beq
g_{\mu\nu} \, \rightarrow \,g_{\mu\nu} + h_{\mu\nu}\,,
\qquad
h_{0\mu}=0\,,
\qquad
\pa_{i}\,h^{ij}=0
\qquad
\mbox{and}
\qquad
h_{ii}=0\,,
\label{GW}
\eeq
where the last three conditions mean synchronous coordinate
condition and fixing the gauge freedom such that we deal with
the tensor mode only. The background metric should be
\ $g_{\mu\nu}^{0}\,=\,\{1,\,\,-\delta_{ij}\,a^2(t)\}$,
where $a(t)$ can be chosen as cosmological constant-, radiation-
or dust-dominated classical solution. Finally, our notations
are $\mu\,=\,0,i$  \ and \ $i\,=\,1,2,3$. In order to explore
the time dynamics of the gravitational waves one can make a
partial Fourier transformation
\beq
h_{\mu\nu}(t,{\vec r})\,=\,\int \frac{d^3k}{(2\pi)^3}\,
e^{i{\vec r}\cdot{\vec k}}\,h_{\mu\nu}(t,{\vec k})
\label{Furier}
\eeq
and assume that the modes with different momenta do not
interact between each other. Then $k=|{\vec k}|$ becomes
a constant parameter and one can deal with an ordinary
differential equation instead of a partial one.

In the original papers we worked with both classical case described
only by the action (\ref{vacuum}) in Ref.~\refcite{HD-Stab} and
with the theory which includes semiclassical corrections
(\ref{induc}) in Ref.~\refcite{GW-Stab}. It was shown that the
effect of these semiclassical corrections is negligible
when we deal with the sufficiently small perturbations and
sufficiently weak background. The qualitative explanation
of this fact is that all the terms in (\ref{induc}) are at
least of the third order in curvature tensor, or reduce to
the less relevant $R^2$-term. Therefore, for the reason of
compactness we will restrict ourselves by the purely classical
case and also keep the cosmological constant zero and the
space section of the space-time manifold plane.
Then the equation for the perturbations have the form
\beq
&&
\frac{1}{3}\, \stackrel{....}{h}
\,+\, 2H \stackrel{...}{h}
+ \Big(H^2 + \frac{{M_P}^{2}}{32\pi a_1}\Big)\,\ddot{h}
+ \frac{2}{3} \Big(\frac14\,\frac{\na^4 h}{a^{4}}
- \frac{\nabla^2 \ddot{h}}{a^2}
-  H\,\frac{\nabla^{2} \dot{h}}{a^{2}}\Big)
\nonumber
\\
&-&
\Big[ H \dot{H} + \ddot{H} + 6 H^3
\,-\,\frac{{3M_P}^2\,H}{32\pi a_1}\Big] \,\stackrel{.}{h}
\,-\,\Big[\frac{{M_P}^{2}}{32\pi a_1}
- \frac43\, \big(\dot{H} + 2H^2\big)
\Big]\,\frac{\nabla^{2} h}{a^{2}}
\nonumber
\\
&-&
\Big[
\Big(24 \dot{H} H^{2} + 12 \dot{H}^{2} + 16 H \ddot{H}
+ \frac83 \stackrel{...}{H}\Big)
-\frac{{M_P}^{2}}{16\pi a_1}\big(2 \dot{H} + 3 H^{2}\big)\Big]\,h
\,=\,0\,.
\label{pertu}
\eeq
Already at this level one can see that the equation depends only
on the coefficient of the Weyl-squared term $a_1$ in the action
(\ref{HD}) and not on other terms, as one should expect.

The analysis of the equation (\ref{pertu}) and its semiclassical
generalization has been done in Ref.~\refcite{HD-Stab} and Ref.~\refcite{GW-Stab},
correspondingly. Let us present here only qualitative results,
which were achieved by both analytical and numerical methods.
The analytical method was based on the following idea.
One can approximately treat all coefficients as constants,
assuming that the time variation of the Hubble parameter and
its derivatives performs slower that the one of the
perturbations. In this case the consideration can be
performed by conventional elementary methods.
The numerical methods included the CMBEasy software or
Wolfram's Mathematica, and provided the results which
were perfectly consistent with the mentioned analytic
approach.

The net result is that the stability is completely defined
by the sign of the coefficient \ \  $a_1$
of the Weyl-squared term\footnote{Let us mention that the
same is true in the semiclassical case with the non-zero
coefficient $c$, which corresponds to the classical $a_4$.}.
The most relevant observation is that the sign of this term
defines whether graviton or ghost has positive or negative
kinetic energy!

\begin{figure}
\includegraphics[height= 6.0 cm,width=10.0cm]{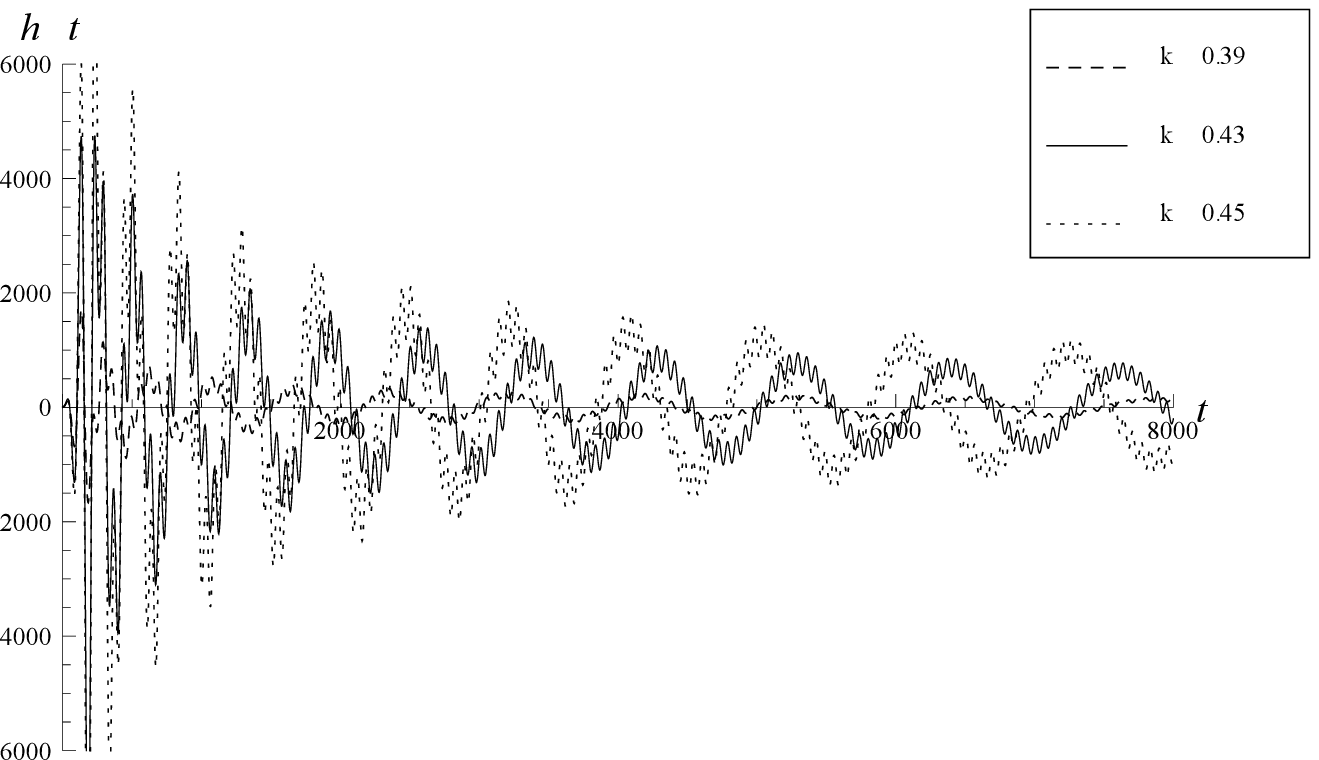}
\includegraphics[height= 6.0 cm,width=10.0cm]{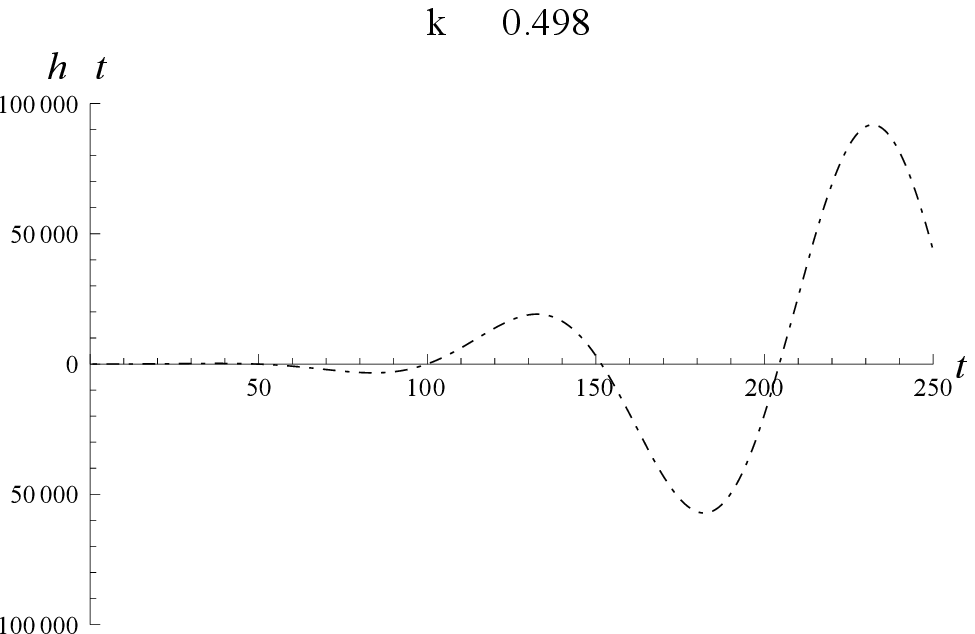}
\caption{Illustrative plots for the case of radiation-dominated
Universe. There are no growing modes up to the frequency
$k \approx 0.5$  in Planck units. Starting from this value, one
can observe the massive ghost making destructive work.}
\label{fig_radiation}
\end{figure}

One can distinguish the following two cases:

\noindent
$\bullet$
\
The coefficient of the Weyl-squared term is negative, $a_1<0$.
\ Then
\beq
G_{\rm spin-2}(k)\,\sim\,\frac{1}{m^2}\,\,\left(
\frac{1}{k^2} - \frac{1}{k^2 + m^2} \right),\quad
m \propto M_P\,.
\label{a1nega}
\eeq
In this case there are no growing modes up to the Planck scale,
${\vec k}^2 \approx M_P^2$.
For the  dS background this is in a perfect agreement with
the previous results of Ref.~\refcite{star83} and Ref.~\refcite{HHR}.
It is remarkable that when the frequency $k=|{\vec k}|$ is getting
close to the Planck scale, the growing modes start to show up.
From the physical side this means that the higher derivative
theory (\ref{vacuum}) is actually stable against ghost-induced
perturbations, but only for the frequencies below the Planck
cut-off. Some plots illustrating this situation are shown
in Figure \ref{fig_radiation}.
\vskip 1mm

\noindent
$\bullet$
\
The classical coefficient of the Weyl-squared term is
positive, $a_1>0$. In this case the propagator of the tensor
mode has the form
\beq
G_{\rm spin-2}(k)\,\sim\,\frac{1}{m^2}\,\,\left(
 \,-\, \frac{1}{k^2} \,+\, \frac{1}{k^2+m^2}\right),\quad
m \propto M_P\,.
\label{a1posi}
\eeq
With this ``wrong'' sign of $a_1$, the massless graviton is
becoming a ghost. On the contrary, massive spin-2 particle
in this case has positive energy. As one could expect,
in this case there is no Planck-mass threshold and, as we
have found, there are rapidly growing modes at any scale of
frequencies. This example is artificial, but very illustrative,
for it explicitly shows the relation between mass of the ghosts
and the stability of classical solutions.

Coming back to the physical case $a_1<0$, the natural interpretation
of the result is that, at low energies, the massive ghosts are present
only in the vacuum state. There are no even one of such excitations
``alive'' until the typical energy scale remains below the Planck
mass threshold. As far as the frequency comes close to $M_P$, the
ghosts start to be generated from vacuum and we observe strong
instabilities.

One more verification of the correspondence between ghosts and
stability with respect to tensor modes can be obtained for the
superrenormalizable theory (\ref{highde}). According to the
result (\ref{propa}), if we include just one d’Alembert
operator $\Box$, there will
be only one extra massive tensor particle and it is not a ghost,
hence the stability conditions should not be modified. And this
is exactly what we have observed in this case by means of
numerical and analytical methods, in Ref. ~\refcite{HD-Stab}.
In particular, the threshold frequency almost does not change
if the mass of the ghost is the same. Once again, we observe a
correspondence between the presence and mass of the ghosts and
(in)stability of the classical solution.

\section{Conclusions and further perspectives}

One should definitely quantize both matter and gravity,
for otherwise the theory would not be complete. Indeed, the
quantum matter is something essentially more certain than the
quantum metric, simply because quantization of matter is really
experimentally supported, in all senses. Since it is not
possible to deal with the quantization of matter fields without
higher derivatives terms (\ref{HD}), the main question is not
whether we like these terms or not. In our opinion the question
is to explain why these terms do not produce destructive
instabilities in the classical gravitational solutions.

For QG with higher derivatives, the propagator includes
massive nonphysical mode(s) called ghosts. These massive
ghosts are capable to produce terrible instabilities, but
for some unknown reason our world is stable and it is
interesting to understand why this happens. There are many
ways to deal with this fact, and here we reviewed some of
them. It turns out that, at least in the cosmological case,
ghosts are not produced at the energy scales far below the
Planck mass. If there is no at least one such ghost excitation
in the initial spectrum, there are no instabilities at the
linear approximation and the Lyapunov theorems guarantee that
this will be the case, also, at the non-linear level.

It is possible that  massive ghosts do not pose real danger
below the Planck scale. However, in order to check this fact,
it is important to have well-established results for other
backgrounds, starting from the Schwarzschild and Kerr
solutions. On the other hand, it would be interesting to
analyse the general metric case in the approximation of
weak curvature tensor components.

Assuming that our conjecture about the situation with ghosts
 ``sleeping'' in the vacuum state is correct, the higher derivative
gravity becomes  a perfect candidate to be an effective QG below
the Planck scale. Then we have to answer the question of whether
the effect of this theory at low energies is the same of the
low-energy quantum GR or not, as it was discussed in Ref.~\refcite{Polemic}.

It is clear that the energy scale below Planck mass covers most of
the possible applications. On the other hand, there is a conceptually
important question of what happens with the ghosts above $M_P$. In
this case we need some new ideas. The solution can be related to
string theory, or to some new principles of Physics which we do
not know yet. In principle, on of the options would be some
principle which forbids the Planck densities of energy to form.
For instance, some hypothesis which closely fit this requirements,
can be found in the recent works \cite{DG}, but, in general, this
problem remains open.

\section*{Acknowledgments}

This short review is based on original papers by the authors and
their collaborators. We are grateful to them, especially to Julio
Fabris, for the contributions to these works. Another source of
this review are seminars, mainly given by I.Sh. between 2012 and
2014. We appreciate the contribution of those who invited him
to speak about the subject and also those who asked questions.
The work of the authors was partially supported by
CNPq, CAPES, FAPEMIG and (in case of I.Sh.) ICTP.


\end{document}